\documentstyle[12pt]{article}

\def\beq{\begin{equation}}
\def\eeq{\end{equation}}
\def\bea{\begin{eqnarray}}
\def\eea{\end{eqnarray}}

\def\ba{\begin{array}}
\def\ea{\end{array}}

\def\,{\"{U}}
\def\6{\.{I}}

\begin{document}
\title{Exact solution of Schr\"{o}dinger equation for Pseudoharmonic
potential}

\author{Ramazan Sever$^1$\thanks{Corresponding Author: sever@metu.edu.tr},
Cevdet Tezcan$^2$, Metin Akta\c{s}$^3$,\"{Ozlem Ye\c{s}ilta\c{s}$^4$} \\[1cm]
$^{1}$Department of Physics, Middle East Technical University \\
06531 Ankara, Turkey \\[.5cm]
$^{2}$Faculty of Engineering, Ba\c{s}kent University, Ba\~{g}l{\i}ca
Campus,\\
Ankara, Turkey\\[.5cm]
$^{3}$Department of Physics, Faculty of Arts and Sciences,
K{\i}r{\i}kkale University,\\
K{\i}r{\i}kkale, Turkey,\\
 $^4$Turkish Atomic Energy Authority, Istanbul Road 30km, Kazan, Ankara, Turkiye}

\date{\today}
\maketitle
\normalsize

\begin{abstract}
Exact solution of Schr\"{o}dinger equation for the pseudoharmonic
potential is obtained for an arbitrary angular momentum. The energy
eigenvalues and corresponding eigenfunctions are calculated by
Nikiforov-Uvarov method. Wavefunctions are expressed in terms of
Jacobi polynomials. The energy eigenvalues are calculated
numerically for some values of $\ell$ and $n$ with $n\leq 5$ for
some diatomic molecules.
\end{abstract}


\date{\today}
\maketitle
\normalsize

\newpage
\section{Introduction}
The study of  anharmonic oscillators in three dimensions has much
attention in chemical physics phenomena [1-45]. They have wide
applications in molecular physics [1]. The solutions of
Schr\"{o}dinger equation for any $\ell$-state for such potentials
have also a special interest. The Morse potential is commonly used
for an anharmonic oscillator. However, its wavefunction does not
vanish at the origin, but the Mie type and the pseudoharmonic
potentials do. The Mie type potential has the general features of
the true interaction energy [1], interatomic and inter-molecular and
dynamical properties of solids [2]. The pseudoharmonic potential may
be used for the energy spectrum of linear and non-linear systems
[3]. The Mie type and pseudoharmonic are two exactly solvable
potentials other than the Coulombic and harmonic oscillator. This
potential may be considered as an intermediate potential between the
harmonic oscillator potential and anharmonic potentials, as the
Morse oscillator (MO) potential, which is a more realistic
potential, in good agreement with the experimental spectroscopical
data [47]. A comparative analysis of potentials HO-3D (3-dimensional
harmonic oscillator potential) and PHO is performed in [2].

To solve the Schr\"{o}dinger equation for the pseudoharmonic
potential, the Nikiforov-Uvarov method [46] which is introduced for
the solution of the hypergeometric type second order differential
equations appeared in the time-independent problems. The method is
based on the determination of the solution in terms of special
orthogonal functions for any general second-order differential
equations [47].

By taking an appropriate $s=s(r)$ coordinate transformation, the
Schr\"{o}dinger equation in spherical coordinates is reduced to a
generalized equation of hypergeometric type as
\begin{equation}
\Psi^{\prime\prime}(s)+\frac{\tilde{\tau}(s)}{\sigma(s)}\Psi^{\prime}(s)+\frac{\tilde{\sigma}(s)}{\sigma^{2}(s)}\Psi(s)=0,\label{eq1}
\end{equation}
where, $\sigma(s)$ and $\tilde{\sigma}(s)$ are polynomials, at most
second-degree, and $\tilde{\tau}(s)$ is in a first-degree
polynomial.

The contents of this work is as follows. In section 2, solution of
Schr\"{o}dinger equation for any $\ell-$state is introduced. In
section 3, solutions for Mie-type and pseudoharmonic potentials are
given. As an example, some numerical values of the energy levels of
$N_{2}$, $CO$, $NO$ and $NH$ molecules are computed. Section 4 is
devoted for the calculations.

\section{Method and Solutions}
The pseudoharmonic potential has the form [3]

\begin{equation}
V(r)=V_{0}\left(\frac{r}{r_{0}}-\frac{r_{0}}{r}\right)^{2},\label{eq2}
\end{equation}
where $V_{0}$ is the dissociation energy and $r_{0}$ is the
equilibrium intermolecular separation. The radial part of the
Schr\"{o}dinger equation for a diatomic molecule potential is

\begin{equation}
\left(-\frac{\hbar^{2}}{2\mu}\frac{1}{r^{2}}\frac{d}{dr}\left(r^{2}\frac{d}{dr}\right)+\frac{\ell(\ell+1)\hbar^{2}}{2\mu
r^{2}}+V(r)\right)R_{n\ell}(r)=E_{n\ell}R_{n\ell}(r),\label{eq3}
\end{equation}
where $\mu$ is the reduced mass of the diatomic molecules. $n$
denotes the radial quantum number ($n$ and $\ell$ are named as the
vibration-rotation quantum numbers in molecular chemistry), $r$ is
the internuclear separation. Substituting the explicit form of
$V(r)$, we get

\begin{equation}
\frac{d^{2}R_{n\ell}(r)}{dr^{2}}+\frac{2}{r}\frac{dR_{n\ell}(r)}{dr}
+\frac{2\mu}{\hbar^{2}}\left(E_{n\ell}-V_{0}\left(\frac{r}{r_{0}}-\frac{r_{0}}{r}\right)^{2}-\frac{\ell(\ell+1)\hbar^{2}}{2\mu
r^{2}}\right)R_{n\ell}(r)=0.\label{eq4}
\end{equation}
By defining a coordinate transformation

\begin{equation}
r^{2}=s,\label{eq5}
\end{equation}
and the following dimensionless parameters

\begin{eqnarray}
\alpha^{2}=\frac{V_{0}}{r_{0}^{2}}\frac{\mu}{2\hbar^{2}}
\label{eq6}\\[0.2cm]
\varepsilon=\left(E_{n\ell}+2V_{0}\right)\frac{\mu}{2\hbar^{2}}
\label{eq7}\\[0.2cm]
\beta=\left(V_{0}r_{0}^{2}+\frac{\ell(\ell+1)\hbar^{2}}{2\mu}\right)\frac{\mu}{2\hbar^{2}},\label{eq8}
\end{eqnarray}
we obtain

\begin{equation}
\frac{d^{2}R}{ds^{2}}+\frac{3/2}{s}\frac{dR}{ds}+\frac{1}{s^{2}}\left(-\alpha^{2}s^{2}+\varepsilon
s-\beta\right)R(s)=0.\label{eq9}
\end{equation}
Thus, comparing the Eqs. (1) and (8), we obtain the following
polynomials

\begin{equation}
\tilde{\tau}(s)=\frac{3}{2},\quad \sigma(s)=s,\quad and \quad
\tilde{\sigma}(s)=-\alpha^{2}s^{2}+\varepsilon s-\beta.\label{eq10}
\end{equation}
To find a particular solution of Eq. (9), we apply separation
variables as $R_{n\ell}(s)=\phi(s)y(s)$, Schr\"{o}dinger equation,
Eq. (1) is reduced to an equation of hypergeometric type

\begin{equation}
\sigma(s)y^{\prime\prime}+\tau(s)y^{\prime}+\lambda y=0,\label{eq11}
\end{equation}
where $\phi(s)$ is a solution of
$\phi^{\prime}(s)\neq\phi(s)=\pi(s)/\sigma(s)$ and $y(s)$ is a
hypergeometric type function whose polynomial solutions are given by
the well-known Rodrigues relation [46]

\begin{equation}
y(s)=y_{n}(s)=\frac{B_{n}}{\rho(s)}\frac{d^{n}}{ds^{n}}\left[\sigma^{n}(s)\rho(s)\right].\label{eq12}
\end{equation}
Here, $\sigma(s)$ is a polynomial of degree at most one, $\rho(s)$
is a weight function and $B_{n}$ is a normalization constant. It
satisfy the condition

\begin{equation}
(\sigma\rho)^{\prime}=\tau\rho.\label{eq13}
\end{equation}
The function $\pi$ and parameter $\lambda$ are defined as

\begin{equation}
\pi=\left(\frac{\sigma^{\prime}-\tilde{\tau}}{2}\right)\pm\sqrt{\left(\frac{\sigma^{\prime}-\tilde{\tau}}{2}\right)^{2}
-\tilde{\sigma}+k\sigma},\label{eq14}
\end{equation}
and

\begin{equation}
\lambda=k+\pi^{\prime}.\label{eq15}
\end{equation}
Here $\pi(s)$ is a polynomial with the parameter $s$. Determination
of $k$ is the essential point in the calculation of $\pi(s)$. $k$ is
determined by setting the square root as the square of a polynomial.
Thus a new eigenvalue equation for the Schr\"{o}dinger equation
becomes

\begin{equation}
\lambda=\lambda_{n}=-n\tau^{\prime}-\frac{n(n-1)}{2}\sigma^{\prime\prime};\quad
n=0,1,2\ldots,\label{eq16}
\end{equation}
where

\begin{equation}
\tau(s)=\tilde{\tau}(s)+2\pi(s),\label{eq17}
\end{equation}
and it will have a negative derivative. Therefore the polynomial of
$\pi(s)$ is found in four positive values

\begin{eqnarray}
\pi(s) &= -\frac{1}{4}\pm\left(\alpha s+\sqrt{\beta+\frac{1}{16}}\right),\quad k_{+}=\varepsilon+\gamma \nonumber\\
      &=-\frac{1}{4}\pm\left(\alpha s-\sqrt{\beta+\frac{1}{16}}\right),\quad k_{-}=\varepsilon-\gamma,\label{eq18}
\end{eqnarray}
where $\gamma=2\alpha\sqrt{\beta+\frac{1}{16}}$. By following Eq.
(17), we get

\begin{eqnarray}
\tau(s)=\frac{3}{2}+2\left\{\begin{array}{ll}
-\frac{1}{4}\pm\left(\alpha s+\sqrt{\beta+\frac{1}{16}}\right),
~~&\mbox{$for~~k_{+}=\varepsilon+\gamma$}\\[0.5cm]
-\frac{1}{4}\pm\left(\alpha s-\sqrt{\beta+\frac{1}{16}}\right),
~~&\mbox{$for~~k_{-}=\varepsilon-\gamma$}.
\end{array}
\right.
\end{eqnarray}
To have a negative derivative of $\tau(s)$ and a physical
eigenfunction, we take the following substitution: for {\it
$k_{-}=\varepsilon-\gamma$}, we have

\begin{equation}
\pi(s)=-\frac{1}{4}-\left(\alpha
s-\sqrt{\beta+\frac{1}{16}}\right),\label{eq29}
\end{equation}
and

\begin{equation}
\tau(s)=1-2\left(\alpha
s-\sqrt{\beta+\frac{1}{16}}\right),\label{eq30}
\end{equation}
Thus, using Eq.(16) we obtain

\begin{equation}
\lambda=\lambda_{n}=2n\alpha,\label{eq31}
\end{equation}
and using Eqs.(22) and (15) we obtain the following energy
eigenvalues,

\begin{equation}
\varepsilon=\left(2n+1+2\sqrt{\beta+\frac{1}{16}}\right)\alpha.\label{eq32}
\end{equation}
Substituting Eq.(23) into Eq. (7), the energy eigenvalue becomes

\begin{equation}
E_{n\ell}=-2V_{0}+\frac{\hbar}{r_{0}}\sqrt{\frac{2V_{0}}{\mu}}\left[(2n+1)+2\sqrt{\frac{\mu}{2\hbar^{2}}
\left(V_{0}r_{0}^{2}+\frac{\ell(\ell+1)\hbar^{2}}{2\mu}\right)
+\frac{1}{16}}\right].\label{eq33}
\end{equation}
In order to calculate the wave function, the weight function $\rho$
can be obtained by using Eq.(13) as

\begin{equation}
\rho(s)=s^{2\sqrt{\beta+\frac{1}{16}}}~e^{-2\alpha s},\label{eq34}
\end{equation}
and using Eq.(12) which is Rodrigues relation, we can get the
function $y$,

\begin{equation}
y_{n}(s)=s^{1+2\sqrt{\beta+\frac{1}{16}}}~e^{-2\alpha s}\label{eq35}
\end{equation}
Finally  the wave function is obtained as

\begin{equation}
R_{n\ell}=s^{-\frac{1}{4}+\sqrt{\beta+\frac{1}{16}}}~e^{-\alpha
s}~L_{n}^{2\sqrt{\beta+\frac{1}{16}}}(2\alpha s).\label{eq36}
\end{equation}

\section{Conclusions}

We have studied analytical solution of Schr\"{o}dinger equation for
a Pseudoharmonic potential.  Energy eigenvalues and the
corresponding wave functions are calculated for a diatomic system
with any angular momentum $L$ . The NU method is used in the
computations. Numerical values of energy for $N_{2}$, $CO$, $NO$ and
$NH$ molecules are calculated for different values of principal and
angular quantum numbers $n$ and $\ell$. Results are listed in Table
I. The potential parameters are obtained from Ref. [10].

\section{Acknowledgements}

This research was partially supported by the Scientific and
Technological Research Council of Turkey.

\newpage
\begin{table}[tbp]
\caption{Energy eigenvalues $(in~eV)$ of pseudoharmonic potential
for $N_2$, $CO$, $NO$ and $CH$ diatomic molecules with different
values of $n$ and $\ell$. Necessary parameters are given in [10].}

\begin{tabular}{cccccc}\hline\hline\\
$n$&$\ell$&$N_2$&$CO$&$NO$&$CH$\\[0.5cm]\hline
0&0&0.10915590&0.10193061&0.08248827&0.16863440\\[0.2cm]\hline
1&0&0.32734304&0.30567217&0.24735916&0.50500718\\[0.2cm]
 &1&0.32784167&0.30615078&0.24778171&0.50859034\\[0.2cm]\hline
2&0&0.54553018&0.50941373&0.41223005&0.84137996\\[0.2cm]
 &1&0.54602881&0.50989234&0.41265260&0.84496312\\[0.2cm]
 &2&0.54702603&0.51084953&0.41349768&0.85212458\\[0.2cm]\hline
4&0&0.98190446&0.91689685&0.74197183&1.51412550\\[0.2cm]
 &1&0.98240309&0.91737546&0.74239438&1.51770870\\[0.2cm]
 &2&0.98340031&0.91833265&0.74323946&1.52487010\\[0.2cm]
 &3&0.98489606&0.91976835&0.74450700&1.53560020\\[0.2cm]
 &4&0.98689026&0.92168247&0.74619689&1.54988430\\[0.2cm]\hline
5&0&1.20009160&1.12063840&0.90684272&1.85049830\\[0.2cm]
 &1&1.20059020&1.12111700&0.90726527&1.85408150\\[0.2cm]
 &2&1.20158750&1.12207420&0.90811035&1.86124290\\[0.2cm]
 &3&1.20308320&1.12350990&0.90937789&1.87197290\\[0.2cm]
 &4&1.20507740&1.12542400&0.91106778&1.88625710\\[0.2cm]
 &5&1.20756990&1.12781650&0.91317990&1.90407610\\[0.2cm]\hline
\end{tabular}
\end{table}

\newpage

\begin{thebibliography}{10}

\bibitem{ref38} G.C. Maitland, M. Rigby, E. B. Smith and W. A. Wakeham, {\it \.{I}ntermolecular forces}
(Oxford Univ. Press, Oxford, 1987)
\bibitem{ref39} M. L. Klein and J. A. Vemebles, {\it Rare gas solids, Vol.1}
(Academic Press, New York, 1976)
\bibitem{ref40} I. I. Goldman and V. D. Krivchenkov, {\it Problems in quantum mechanics}
(Pergamon Press, New York, 1961); Y. Weissman and J. Jortner, Phys.
Lett. A {\bf 70} 177 (1979); M. L. Szego, Chem. Phys. {\bf 87} 431
(1984); M. Sato and J. Goodisman, Am. J. Phys. {\bf 53} 350 (1985);
\c{S}. Erko\c{c}, R. Sever, Phys. Rev. A {\bf 37} 2687 (1988).
\bibitem{ref43} By Julian, J. I. Ting, J. Phys. B {\bf 11} 1 (1994).
\bibitem{ref44} J. Plieva, J. Mol. Spectrosc. {\bf 193} 7 (1997).
\bibitem{ref45} M. Karplus, R. N. Porter, {\it Atoms and Molecules: An Introduction For Students of Physical Chemistry}
(Benjamin, Menlo Park, CA., 1970).
\bibitem{ref46} G. Herzberg, {\it Spectra of Diatomic Molecules}
(Van Nostrand, Princeton, NJ., 1950).
\bibitem{ref47} C. Berkdemir, A. Berkdemir, J. Han, Chem. Phys.
Lett. {\bf 417} 326 (2006).

%%%%%%%%%%%%%%%%%%%%%%%%%%%%%%%%%%%%%%%%%%%%%%%%%%%%%%%%%%%%%%%%%%%%%%%%%%%%%%%%%%%%%%%%%%%

\bibitem{ref1} A. J. Sous, Mod. Phys. Lett. {\bf A21}, 1675 (2006).
\bibitem{ref2} S. H. Dong and M. Lozada-Cassou, Int. J. Mod. Phys. {\bf B19}, 4219
(2005); Phys. Lett. {\bf A340}, 94 (2005).
\bibitem{ref3} S. H. Dong, Y. Tang, G. H. Sun et al., Ann. Phys. (NY) {\bf 315}, 566 (2005).
\bibitem{ref4} J. A. Yu and S. H. Dong, Phys. Lett. {\bf A325}, 194 (2004).
\bibitem{ref5} G. Chen, Z. D. Chen, Z. M. Lou, Chinese Phys. {\bf 13}, 279 (2004).
\bibitem{ref6} N. Saad, R. L. Hall and H. \c{C}ift\c{c}i, J. Phys. A: Math. Gen. {\bf 39}, 8477
(2006).
\bibitem{ref7} A. Rajaki, Commun. Theor. Phys. {\bf 45}, 609 (2006).
\bibitem{ref8} T. Barakat, Phys. Lett. {\bf A344}, 411 (2005).
\bibitem{ref9} A. Turbiner, Lett. Math. Phys. {\bf 74}, 169 (2005).
\bibitem{ref10} M. C. Zhang and Z. B. Wang, Chinese Phys.
Lett. {\bf 22}, 2994 (2006).
\bibitem{ref11} H. A. Alhendi, E. \.{I}. Lashin,  Can. J. Phys. {\bf 83}, 541 (2005).
\bibitem{ref12} B. P. Mahapatra, Int. J. Mod. Phys. {\bf A20}, 2687
(2005).
\bibitem{ref13} S. K. Maayedi, M. Solimannejad and A. F. Jalbout, Int. J. Quant Chem. {\bf 100}, 16 (2004).
\bibitem{ref14} F. L. Lu and C. Y. Chen, Acta Phys. Sin-Chin.ed. {\bf 53}, 688 (2004).
\bibitem{ref15} D. Chelkak, P. Kargaev and E. Korotyaev, Lett. Math. Phys. {\bf 64}, 7 (2003).
\bibitem{ref16} M. Znojil, D. Yanovich and V. P. Gerdt, J. Phys. {\bf A36},
6531 (2003).
\bibitem{ref17} S. H. Dong, Int. J. Theor. Phys. {\bf 41}, 1991 (2002).
\bibitem{ref18} S. H. Dong, Phys. Scripta {\bf 65}, 289 (2002).
\bibitem{ref19} M. Jafapour and D. Afshar, J. Phys. {\bf A35}, 87 (2002).
\bibitem{ref20} S. H. Dong, Int. J. Theor. Phys. {\bf 41}, 89 (2002).

%%%%%%%%%%%%%%%%%%%%%%%%%%%%%%%%%%%%%%%%%%%%%%%%%%%%%%%%%%%%%%%%%%%%%%%%%%%%%%%%%%%%%%%%%%%

\bibitem{ref21} S. Bose and N. Gupta, \.{I}L Nuovo \c{C}imento {\bf B114}, 299 (1998).
\bibitem{ref22} A. Khare, S. N. Behra, Pramana J. Phys. {\bf 14},
327 (1980).
\bibitem{ref23} D. Emin, Phys. Today {\bf 35}, 35 (1982); D. Emin and T. Holstein, Phys. Rev. Lett. {\bf 36}, 323 (1976).
\bibitem{ref24} S. Coleman, {\it Aspect of Symmetry}, selected Erice
Lectures (Cambridge Univ. Press, Cambridge, 1988, p. 234).
\bibitem{ref25} R. S. Kaushal, Ann. Phys. (NY) {\bf 206}, 90 (1991);
Phys. Lett. {\bf A142}, 57 (1989).
\bibitem{ref26} R. S. Kaushal and D. Parashar, Phys. Lett. {\bf A170},
335 (1992).
\bibitem{ref27} S. \"{O}z\c{c}elik, M. \c{S}im\c{s}ek, Phys. Lett. {\bf A152}, 145
(1991); M. \c{S}im\c{s}ek, S. \"{O}z\c{c}elik, Il Nuovo cimento. B
(Nuovo cimento, B) \textbf{114(1)} (1999) 87-92.
\bibitem{ref28} S. H. Dong, Z. Q. Ma and G. Esposito, Found Phys. Lett. {\bf 12}, 465 (1999).
\bibitem{ref29} S. H. Dong, Z. Q. Ma, J. Phys. {\bf A31}, 9855 (1998).
\bibitem{ref30} S. H. Dong, Int. J. Theor. Phys. {\bf 39}, 1119
(2000); {\bf 40}, 569 (2001); {\bf 41}, 89 (2002).
\bibitem{ref31} M. S. Child, S. H. Dong and X. G. Wang, J. Phys. {\bf A33}, 5653 (2000).
\bibitem{ref32} S. H. Dong, Physica Scripta {\bf 64}, 273
(2001).
\bibitem{ref33} M. Znojil, J. Math. Phys. {\bf 30}, 23 (1989); {\bf 31}, 108 (1990).
\bibitem{ref34} A. Voros, J. Phys. {\bf A32}, 5993 (1999).
\bibitem{ref35} G. Esposito, Found Phys. Lett. {\bf 11}, 535 (1998).
\bibitem{ref36} A. O. Barut, J. Math. Phys. {\bf 21}, 586 (1980).
\bibitem{ref37} G. Esposito, J. Phys. {\bf A31}, 9493 (1998);M. Sage, M. J. Goodisman:
Am. J. Phys. \textbf{53} (1985) 350
\bibitem{ref}   A. F. Nikiforov, V. B. Uvarov, {\it Special functions of Mathematical Physics}
(Birkh\"{a}user, Basel, 1988)
\bibitem{ref}    G. Szego, {\it Orthogonal polynomials}
(American Mathematical Society, New York, 1959), (revised edition)
%%%%%%%%%%%%%%%%%%%%%%%%%%%%%%%%%%%%%%%%%%%%%%%%%%%%%%%%%%%%%%%%%%%%%%%%%%%%%%%%%%%%%%%%%%%
\end{thebibliography}
\end{document}